\documentclass[journal]{IEEEtran}

\usepackage{amsfonts}
\usepackage{amsmath}
\usepackage{amssymb}
\usepackage{lscape}
\usepackage{epsf}
\usepackage{graphicx}
\usepackage{psfrag}

\newtheorem{corollary}{Corollary}
\newtheorem{proposition}{Proposition}
{\indent Claim}
\newtheorem{EXAMPLE}{Example}

\newcommand{\cM}{{\mathcal{M}}}

\newcommand{\cS}{{\mathcal{S}}}

\newcommand{\bldalpha}{{\mbox{\boldmath $\alpha$}}}
\newcommand{\bldsmallalpha}{{\mbox{\scriptsize \boldmath $\alpha$}}}

\newcommand{\bldbeta}{{\mbox{\boldmath $\beta$}}}
\newcommand{\bldsmallbeta}{{\mbox{\scriptsize \boldmath $\beta$}}}

\newcommand{\bldomega}{{\mbox{\boldmath $\omega$}}}
\newcommand{\bldsmallomega}{{\mbox{\scriptsize \boldmath $\omega$}}}

    \def\squarebox#1{\hbox to #1{\hfill\vbox to #1{\vfill}}}

\newlength{\Algwidth}

\title{On a Class of Doubly-Generalized LDPC Codes with Single Parity-Check Variable Nodes}

\author{Enrico Paolini, Mark F. Flanagan, Marco Chiani and Marc P. C. Fossorier
\thanks{E. Paolini and M. Chiani are with DEIS/WiLAB, University of Bologna, via Venezia 52, 47023 Cesena (FC), Cesena, Italy (e-mail: e.paolini@unibo.it, marco.chiani@unibo.it).}%
\thanks{M. F. Flanagan was with Institut f\"ur Mathematik, University of Z\"urich. He is now with the School of Electrical, Electronic and Mechanical Engineering, University College Dublin, Belfield, Dublin 4, Ireland (e-mail: mark.flanagan@ieee.org).}%
\thanks{M. P. C. Fossorier is with ETIS ENSEA, UCP, CNRS UMR-8051, 6 avenue du Ponceau, 95014 Cergy Pontoise, France (e-mail:  mfossorier@ieee.org).}
}

\begin{document}
\maketitle
\thispagestyle{empty}
\pagestyle{empty}

\begin{abstract}
A class of doubly-generalized low-density parity-check (D-GLDPC) codes, where single parity-check (SPC) codes are used as variable nodes (VNs), is investigated. An expression for the growth rate of the weight distribution of any D-GLDPC ensemble with a uniform check node (CN) set is presented at first, together with an analytical technique for its efficient evaluation. These tools are then used for detailed analysis of a case study, namely, a rate-$1/2$ D-GLDPC ensemble where all the CNs are $(7,4)$ Hamming codes and all the VNs are length-$7$ SPC codes. It is illustrated how the VN representations can heavily affect the code properties and how different VN representations can be combined within the same graph to enhance some of the code parameters. The analysis is conducted over the binary erasure channel. Interesting features of the new codes include the capability of achieving a good compromise between waterfall and error floor performance while preserving graphical regularity, and values of threshold outperforming LDPC counterparts.
\end{abstract}

\section{Introduction}

Recently, low-density parity-check (LDPC) codes \cite{gallager63:low-density} have been intensively studied due to their near-Shannon-limit performance under iterative belief-propagation decoding. It is usual to represent an LDPC code as a bipartite graph (known as a Tanner graph \cite{tanner81:recursive}), where the nodes are grouped into two disjoint sets, namely, the variable nodes (VNs) and the check nodes (CNs), such that each edge may only connect a VN with a CN. Here, a degree-$q$ VN can be interpreted as a length-$q$ repetition code, as it repeats $q$ times its single information bit toward the CNs. Similarly, a degree-$s$ CN of an LDPC code can be interpreted as a length-$s$ single parity-check (SPC) code, as it checks the parity of the $s$ VNs connected to it.

Doubly-generalized LDPC (D-GLDPC) codes \cite{wang06:D-GLDPC} (see also the previous work \cite{dolinar2003:small_codes}) generalize the concept of LDPC codes. In a D-GLDPC code, a degree-$s$ CN may in principle be any $(s,h)$ linear block code, $s$ being the code length and $h$ the code dimension. Such a CN accounts for $s-h$ linearly independent parity-check equations. Analogously, a degree-$q$ VN may in principle be any $(q,k)$ linear block code, $q$ being the code length and $k$ the code dimension. Such a VN is associated with $k$ D-GLDPC code bits. It interprets these bits as its local information bits and interfaces to the CN set through its $q$ local code bits. 
A D-GLDPC code is said to be \emph{regular} (or \emph{strongly regular}) if all of its VNs are of the same type and all of its CNs are of the same type and is said to be \emph{irregular} otherwise. We point out that the properties of a D-GLDPC code are heavily affected by the generator matrix used to represent its VNs, i.e., by the association between local input words and local codewords of any VN. On the other hand, the overall code properties do not depend on the representation of its CNs. Therefore, by \emph{type of a VN} we mean its local input-output weight enumerating function (IO-WEF), while by \emph{type of a CN} we mean its local weight enumerating function (WEF). Among irregular D-GLDPC ensembles, we call \emph{weakly regular} any ensemble where all the CNs have the same WEF and where all the VNs have the same WEF but may have a different IO-WEF (i.e., they are associated with the same code, but are represented by a different generator matrix). Note that weakly regular D-GLDPC codes preserve the graphical regularity as all the VNs (resp. CNs) have the same degree.

An analysis of the stability condition over the binary erasure channel (BEC) suggests that single parity-check (SPC) codes used as VNs can offer some benefits when codes with local minimum distance larger than $2$ are employed as CNs \cite{paolini08:stability}. In this paper we elaborate on this idea and propose an analysis of a class of strongly and weakly regular D-GLDPC codes where all the CNs have a local minimum distance larger than $2$ and all the VNs are SPC codes. As proved in \cite{flanagan08:growth}, the absence of CNs with minimum distance $2$ is sufficient to have a growth rate of the weight distribution $G(\alpha)$ (see Section \ref{subsec:growth_rate}) such that $\alpha^* \triangleq \inf \{\alpha>0 | G(\alpha) \geq 0 \}$ is strictly positive, which implies an exponentially small number of codewords of small weight linear in the block length. 

The threshold analysis over the BEC for any irregular D-GLDPC ensemble is reviewed in Section~\ref{subsec:threshold}. Two new results, namely, an expression for the growth rate of the weight distribution of any D-GLDPC ensemble with a uniform CN set (i.e., all the CNs are of the same type), and an efficient means of its evaluation based on a polynomial system, are presented in Section~\ref{subsec:growth_rate} and Section~\ref{sub:efficient_eval}. In Section~\ref{sec:rate_one_half} asymptotic and finite length analyses of a case study are presented. More specifically, strongly and weakly regular rate-$1/2$ D-GLDPC codes, where all the CNs are $(7,4)$ Hamming codes and all the VNs are length-$7$ SPC codes, are investigated. The  $(3,6)$ regular LDPC ensemble is used as a benchmark for the new class of codes, as it offers the best threshold over the BEC among rate-$1/2$ LDPC codes with a regular Tanner graph \cite{shokrollahi05:optimal_regular}.


\section{Preliminaries and Notation}\label{section:irregular_D_GLDPC}
We define a D-GLDPC code ensemble $\cM_n$ as follows, where $n$ denotes the number of VNs. There are $n_c$ different CN types $t \in I_c = \{ 1,2,\cdots, n_c\}$, and $n_v$ different VN types $t \in I_v = \{ 1,2,\cdots, n_v\}$. For each CN type $t \in I_c$, we denote by $h_t$, $s_t$ and $r_t$ the CN dimension, length and minimum distance, respectively. For each VN type $t \in I_v$, we denote by $k_t$, $q_t$ and $p_t$ the VN dimension, length and minimum distance, respectively. For $t \in I_c$, $\rho_t$ denotes the fraction of edges connected to CNs of type $t$. Similarly, for $t \in I_v$, $\lambda_t$ denotes the fraction of edges connected to VNs of type $t$. Note that all of these variables are independent of $n$.

The polynomials $\rho(x)$ and $\lambda(x)$ are defined by $\rho(x) \triangleq \sum_{t\in I_c} \rho_t x^{s_t - 1}$ 
and $\lambda(x) \triangleq \sum_{t \in I_v} \lambda_t x^{q_t - 1}$. If $E$ denotes the number of edges in the Tanner graph, the number of CNs of type $t\in I_c$ is then given by $E \rho_t / s_t$, and the number of VNs of type $t\in I_v$ is then given by $E \lambda_t / q_t$. Denoting as usual $\int_0^1 \rho(x) \, {\rm d} x$ and $\int_0^1 \lambda(x) \, {\rm d} x$ by $\int \rho$ and $\int \lambda$ respectively, we see that the number of edges in the Tanner graph is given by $E =n / \int \lambda$ and the number of CNs is given by $m = E \int \rho$. Therefore, the fraction of CNs of type $t \in I_c$ and the fraction of VNs of type $t \in I_v$ are given by
\begin{equation}\label{eq:delta_t_definition}
\gamma_t = \frac{\rho_t}{s_t \int \rho} \quad \textrm{and} \quad \delta_t = \frac{\lambda_t}{q_t \int \lambda}
\end{equation}
respectively. Also the length of any D-GLDPC codeword in the ensemble is given by 
\begin{equation}
N = \sum_{t \in I_v} \left( \frac{E \lambda_t}{q_t} \right) k_t = \frac{n}{\int \lambda} \sum_{t \in I_v} \frac{\lambda_t k_t}{q_t} \; .
\label{eq:DG_LDPC_codeword_length}
\end{equation}
Note that this is a linear function of $n$. Similarly, the total number of parity-check equations for any D-GLDPC code in the ensemble is given by $M = \frac{m}{\int \rho} \sum_{t \in I_c} \frac{\rho_t (s_t - h_t)}{s_t}$. A member of the ensemble then corresponds to a permutation on the $E$ edges connecting CNs to VNs.

The WEF for CN type $t \in I_c$ is given by $A^{(t)}(z) = 1 + \sum_{u=r_t}^{s_t} A_u^{(t)} z^u$. 
Here $A_u^{(t)} \ge 0$ denotes the number of weight-$u$ codewords for CNs of type $t$. The IO-WEF for VN type $t \in I_v$ is given by $B^{(t)}(x,y) = 1 + \sum_{u=1}^{k_t} \sum_{v=p_t}^{q_t} B_{u,v}^{(t)} x^u y^v$.  
Here $B_{u,v}^{(t)} \ge 0$ denotes the number of weight-$v$ codewords generated by input words of weight $u$ for VNs of type $t$. Also, for each $t \in I_v$, corresponding to the polynomial $B^{(t)}(x,y)$ we denote the set $\cS_t = \{ (i,j) \in \mathbb{Z}^2 \; : \; B^{(t)}_{i,j} > 0 \}$.

The design rate of any D-GLDPC ensemble is given by
\begin{equation}\label{eq:design_rate}
R = 1 - \frac{\sum_{t \in I_c} \rho_t (1 - R_t)}{\sum_{t \in I_v} \lambda_t R_t}
\end{equation}
where for $t \in I_c$ (resp. $t \in I_v$) $R_t$ is the local code rate of a type-$t$ CN (resp. VN). 

Throughout this paper, the notation $e = \exp(1)$ denotes Napier's number, all the logarithms are assumed to have base $e$ and for $0<x<1$ the notation $h(x)=-x \log(x) - (1-x) \log(1-x)$ denotes the binary entropy function. 


\section{Asymptotics}\label{section:asymptotics}
 

\subsection{Asymptotic Threshold over the BEC}\label{subsec:threshold}

An extrinsic information transfer (EXIT) chart \cite{ten-brink04:extrinsic} approach can be used to calculate the threshold over the BEC (denoted by $\epsilon^*$) of any irregular D-GLDPC ensemble. Let $\epsilon$ be the BEC erasure probability and $I_A$ be the average \emph{a priori} information. The EXIT function of a type-$t$ $(q_t,k_t)$ VN is given by
\begin{align*}
I_E^{(t)}(I_A,\epsilon) = 1 & - \frac{1}{q_t} \sum_{j=0}^{q_t-1}\,\sum_{z=0}^{k_t} a_{j,z}^{(t)}\,(1-I_A)^j\,I_A^{q_t-j-1}\,\epsilon^z\,(1-\epsilon)^{k_t-z}
\end{align*}
where $a_{j,z}^{(t)} = (q_t-j) \, \tilde{e}_{q_t-j,k_t-z}^{(t)}-(j+1) \tilde{e}_{q_t-j-1,k_t-z}^{(t)}$ and $\tilde{e}_{g,h}^{(t)}$ is the $(g,h)$-th un-normalized split information function \cite{ten-brink04:extrinsic} for a type-$t$ VN. It is defined as the summation of the ranks over all the submatrices obtained by selecting $g$ columns from the generator matrix $\mathbf{G}_t$ of a type-$t$ VN and $h$ columns from the identity matrix $\mathbf{I}_{k_t}$ (of order $k_t$).

The EXIT function of a type-$t$ $(s_t,h_t)$ CN is given by
\begin{align*}
I_E^{(t)}(I_A) = 1 - \frac{1}{s_t} \sum_{j=0}^{s_t-1} a_j^{(t)} \,(1-I_A)^{j}I_A^{s_t-j-1}
\end{align*}
where $a_j^{(t)} = (s_t-j)\,\tilde{e}_{s_t-j}^{(t)}-(j+1)\tilde{e}_{s_t-j-1}^{(t)}$ and $\tilde{e}_g^{(t)}$ is the $g$-th un-normalized information function for a type-$t$ CN. It is defined as the summation of the ranks over all the submatrices obtained by selecting $g$ columns from $\mathbf{G}_t$.

The EXIT function of the whole VN set is given by $I_{E,V}(I_A,\epsilon) = \sum_{t \in I_v}\lambda_t\,I_{E}^{(t)}(I_A,\epsilon)$, while the EXIT function of the whole CN set is given by $I_{E,C}(I_A) = \sum_{t \in I_c}\rho_t\,I_{E}^{(t)}(I_A)$. We highlight that the threshold depends on the VN representations through the split information functions $\tilde{e}_{g,h}$ of the VNs. On the other hand, the threshold does not depend on the representation of the CNs \cite{paolini08:stability}.


\subsection{Growth Rate of the Weight Distribution}\label{subsec:growth_rate}
The growth rate of the weight distribution (or \emph{spectral shape}) of the irregular D-GLDPC ensemble sequence $\{ \cM_n \}$ is defined by\footnote{Note that using (\ref{eq:DG_LDPC_codeword_length}), we may also define the growth rate with respect to the number of D-GLDPC code bits $N$ as 
$H(\gamma) \triangleq \lim_{N\rightarrow \infty} \frac{1}{N} \log \mathbb{E}_{\cM_n} \left[ N_{\gamma N} \right]$. It is straightforward to show that $H(\gamma) = \frac{G(\gamma y)}{y}$ where $y = \frac{1}{\int \lambda} \sum_{t \in I_v} \frac{\lambda_t k_t}{q_t}$.}
\begin{equation}
G(\alpha) \triangleq \lim_{n\rightarrow \infty} \frac{1}{n} \log \mathbb{E}_{\cM_n} \left[ N_{\alpha n} \right]
\label{eq:growth_rate_result}
\end{equation}
where $\mathbb{E}_{\cM_n}$ denotes the expectation operator over the ensemble $\cM_n$, and $N_{w}$ denotes the number of codewords of weight $w$ of a randomly chosen D-GLDPC code in the ensemble $\cM_n$. 
Note that the argument of the growth rate function $G(\alpha)$ is equal to the ratio of D-GLDPC codeword weight to the number of VNs; by (\ref{eq:DG_LDPC_codeword_length}), this captures the behavior of codewords linear in the block length, as in \cite{di06:weight} for the LDPC case. Next, we formulate techniques for evaluation of the growth rate for a D-GLDPC ensemble $\cM_n$ with a uniform CN set, over a wider range of $\alpha$ than was considered in \cite{flanagan08:growth} (where the case $\alpha \rightarrow 0$ was analyzed).

\medskip
\begin{proposition}\label{proposition:growth_rate}
Consider a D-GLDPC ensemble with a uniform CN set. Let $A(z)$ be the WEF of each CN and $B^{(t)}(x,y)$ be the IO-WEF of any type-$t$ VN with $t \in I_v$. The growth rate of the weight distribution is given by
\begin{align}\label{eq:numerical_evaluation_1_weak}
G(\alpha) = \max_{\bldsmallalpha, \bldsmallbeta} \left( \sum_{t \in I_v} X_t^{(\delta_t)}(\alpha_t, \beta_t) + Y(\beta) \right)
\end{align}
where $\bldalpha \triangleq (\alpha_t)_{t \in I_v}$, $\bldbeta \triangleq (\beta_t)_{t \in I_v}$, $\beta \triangleq \sum_{t \in I_v} \beta_t$ and the maximization is subject to the constraints $\alpha_t \ge 0$, $m^{(t)}(\alpha_t) \le \beta \le M^{(t)}(\alpha_t)$ for all $t \in I_v$ and 
\begin{equation}\label{eq:sum_alpha_constraint}
R(\bldalpha, \bldbeta) \triangleq \sum_{t \in I_v} \alpha_t = \alpha \; .
\end{equation}
The expression of $Y(\beta)$ in \eqref{eq:numerical_evaluation_1_weak} is 
$$Y(\beta) = \log \left( \frac{A(z_0)^{\frac{\int \rho}{\int \lambda}}}{z_0^{\beta}} \right) - \frac{h(\beta \int \lambda)}{\int \lambda}
$$
where $z_0$ is the unique positive real solution to
\begin{equation}\label{eq:z0_eqn}
\frac{A'(z_0)}{A(z_0)} \cdot z_0 = \beta \left( \frac{\int \lambda}{\int \rho} \right)
\end{equation}
while (for each $t \in I_v$) the expression of $X_t^{(\delta_t)}(\alpha_t, \beta_t)$ is
$$X_t^{(\delta_t)}(\alpha_t, \beta_t) = \log \left( \frac{ \left( B^{(t)}(x_{0,t},y_{0,t}) \right) ^{\delta_t}}{x_{0,t}^{\alpha_t}y_{0,t}^{\beta_t}} \right)$$  
where $\delta_t$ is defined in \eqref{eq:delta_t_definition} and $(x_{0,t}, y_{0,t})$ are the unique positive real solutions to the pair of simultaneous equations\footnote{The uniqueness of $z_0$, and of $x_{0,t}$ and $y_{0,t}$ for each $t \in I_v$, is guaranteed by Hayman's formula (see for example \cite[Appendix~II]{di06:weight}).}
\begin{equation}
\label{eq:x0t_y0t_eqn_1}
\frac{\partial B^{(t)}}{\partial x}(x_{0,t},y_{0,t}) \cdot \frac{x_{0,t}}{B^{(t)}(x_{0,t},y_{0,t})} = \frac{\alpha_t}{\delta_t}
\end{equation}
\begin{equation}
\label{eq:x0t_y0t_eqn_2}
\frac{\partial B^{(t)}}{\partial y} (x_{0,t},y_{0,t}) \cdot \frac{y_{0,t}}{B^{(t)}(x_{0,t},y_{0,t})} = \frac{\beta_t}{\delta_t} \; . 
\end{equation}
Finally, letting $\bldomega = (\omega_1 \; \omega_2 \; \cdots \; \omega_{k_t})$, we define the function $m^{(t)}(\alpha) = \max_{\bldsmallomega} \sum_{i=1}^{k_t} V_i^{(t)} \omega_i$ where $V_i^{(t)}$ denotes the maximum local codeword weight associated with a local input weight $i \in \{1,2,\cdots,k_t \}$ for a type-$t$ VN (i.e., the maximum $j$ with $(i,j) \in \cS_t$), and the maximization is subject to the constraints $\omega_i \ge 0$ for all $i=1,2,\cdots,k_t$, $\sum_{i=1}^{k_t} \omega_i \le 1$ and $\sum_{i=1}^{k_t} i \omega_i = \alpha$. Also the function $M^{(t)}$ is defined as $M^{(t)}(\alpha) =\min_{\bldsmallomega} \sum_{i=1}^{k_t} U_i^{(t)} \omega_i$ where $U_i^{(t)}$ denotes the minimum local codeword weight associated with a local input weight $i \in \{1,2,\cdots,k_t \}$ for a type-$t$ VN (i.e., the minimum $j$ with $(i,j) \in \cS_t$), and the minimization is subject to the constraints $\omega_i \ge 0$ for all $i=1,2,\cdots,k_t$, $\sum_{i=1}^{k_t} \omega_i \le 1$ and $\sum_{i=1}^{k_t} i \omega_i = \alpha$.
\end{proposition}
\medskip

The proof of Proposition~\ref{proposition:growth_rate} is omitted due to space constraints (it can be found in \cite{flanagan09:SPC_VNs}). Note that (\ref{eq:z0_eqn}) provides an implicit definition of $z_0$ as a function of $\beta$. Similarly, for any $t \in I_v$, (\ref{eq:x0t_y0t_eqn_1}) and (\ref{eq:x0t_y0t_eqn_2}) provide implicit definitions of $x_{0,t}$ and $y_{0,t}$ as functions of $\alpha_t$ and $\beta_t$. 

We deduce as a special case the growth rate for a strongly regular ensemble.

\medskip
\begin{corollary}
The growth rate of the weight distribution for a strongly regular D-GLDPC ensemble is given by
\begin{align}\label{eq:numerical_evaluation_1_strong}
G(\alpha) & = \max_{m(\alpha) \le \beta \le M(\alpha)} \Bigg[ \log \left( \frac{B(x_0,y_0)}{x_0^{\alpha} y_0^{\beta}} \right) \nonumber \\
\, & \phantom{------} + \log \left( \frac{A(z_0)^{\frac{\int \rho}{\int \lambda}}}{z_0^{\beta}} \right) - \frac{h(\beta \int \lambda)}{\int \lambda} \Bigg] 
\end{align}
where $x_0$, $y_0$ and $z_0$ are the unique positive real solutions to (\ref{eq:z0_eqn}) together with
\begin{equation}
\label{eq:x0_y0_eqn_1}
\frac{ \frac{\partial B}{\partial x} (x_0,y_0)}{B(x_0,y_0)} \cdot x_0 = \alpha \quad
\textrm{and} \quad \frac{ \frac{\partial B}{\partial y} (x_0,y_0)}{B(x_0,y_0)} \cdot y_0 = \beta \; .
\end{equation}
\end{corollary}

\subsection{Solution via Polynomial System}\label{sub:efficient_eval}
We solved the optimization problem \eqref{eq:numerical_evaluation_1_weak} using Lagrange multipliers. Letting 
$$
S(\bldalpha, \bldbeta) \triangleq \sum_{t \in I_v} X_t^{(\delta_t)}(\alpha_t, \beta_t) + Y(\beta)
$$
and recalling \eqref{eq:sum_alpha_constraint}, at the maximum we must have
$$
\frac{\partial S(\bldalpha, \bldbeta)}{\partial \alpha_t} = \mu \frac{\partial R(\bldalpha, \bldbeta)}{\partial \alpha_t}
$$
for all $t \in I_v$, where $\mu$ is the Lagrange multiplier. After some calculation, this equation simplifies to $\log x_{0,t} = -\mu$ $\forall\, t \in I_v$. We conclude that all of the $\{ x_{0,t} \}$ are equal, and we may write $x_{0,t} = x_0$ for all $t \in I_v$. At the maximum, we must also have
\[
\frac{\partial S(\bldalpha, \bldbeta)}{\partial \beta_t} = \mu \frac{\partial R(\bldalpha, \bldbeta)}{\partial \beta_t}
\]
for all $t \in I_v$. After some calculation, this equation simplifies to $z_0 y_{0,t} \left( \frac{1 - \beta \int \lambda}{\int \lambda} \right) = 1$ $\forall\, t \in I_v$. 
We conclude that all of the $\{ y_{0,t} \}$ are equal, and we may write $y_{0,t} = y_0$ for all $t \in I_v$. Then the latter equation may be written as
\begin{equation}\label{eq:z0_y0_beta_2}
1 + z_0 y_0 = \frac{z_0 y_0}{\beta \int \lambda} \, .
\end{equation}
Thus, for $n_v>1$, the growth rate may be evaluated by solving numerically the $(2n_v+3) \times (2n_v+3)$ system of nonlinear polynomial equations given by (\ref{eq:z0_eqn}), (\ref{eq:x0t_y0t_eqn_1}), (\ref{eq:x0t_y0t_eqn_2}), (\ref{eq:sum_alpha_constraint}) and (\ref{eq:z0_y0_beta_2}). If all VNs are of the same type ($n_v=1$), (\ref{eq:sum_alpha_constraint}) is redundant and (\ref{eq:z0_eqn}), (\ref{eq:x0t_y0t_eqn_1}), (\ref{eq:x0t_y0t_eqn_2}), (\ref{eq:z0_y0_beta_2}) comprise a $4 \times 4$ system for~numerical~solution.


\section{Asymptotic and Finite Length Analysis of a Rate-$1/2$ D-GLDPC Ensemble}\label{sec:rate_one_half}

We consider as a case study a rate-$1/2$ ensemble where all the CNs are $(7,4)$ Hamming codes, and all the VNs are length-$7$ SPC codes. Three representations of the SPC VNs are considered. The first two are the systematic (S) and the cyclic (C) representations. The third one is what we call the \emph{antisystematic} (A) representation, whose $(k \times (k+1))$ generator matrix is obtained from the generator matrix in S form by complementing each bit in the first $k$ columns\footnote{Note that a $(k \times (k+1))$ generator matrix in A form represents a SPC code if and only if the code length $q=k+1$ is odd. For even $k+1$ we obtain a $d_{\min}=1$ code with one codeword of weight $1$.}.

\begin{table}[t]
\begin{center}
\caption{Asymptotic and Finite Length Parameters of Rate-$1/2$ Strongly and Weakly Regular D-GLDPC Codes}\label{tab:codewords}
\begin{tabular}{ccccccc}
\hline \hline 
 & $v$ & $c$ & A & S & C & WR\\
\hline\hline
\multicolumn{7}{c}{\emph{Asymptotic Parameters}}\\
\hline 
$\epsilon^*$ & & & $0.332$ & $0.415$ & $0.450$ & $0.481$\\
\hline
$\alpha^*$ ($\times 10^{-3}$) & & & $11.7$ & $7.2$ & $10.7$ & $9.7$\\
\hline\hline
\multicolumn{7}{c}{\emph{Finite Length Parameters} ($n=500$)}\\
\hline
CWD A & $9$ & $6$ & $\mathbf{21}$ & $17$ & $27$ & $\mathbf{17}$\\
\hline
CWD B & $9$ & $6$ & $33$ & $\mathbf{13}$ & $\mathbf{21}$ & $21$\\
\hline
CWD C & $9$ & $6$ & $27$ & $15$ & $33$ & $23$\\
\hline
CWD D & $9$ & $6$ & $30$ & $14$ & $27$ & $22$\\
\hline
CWD E & $11$ & $7$ & $37$ & $15$ & $29$ & $27$\\
\hline
CWD F & $12$ & $8$ & $27$ & $23$ & $22$ & $23$\\
\hline
$\alpha^* n$ &   &   & $5.85$ & $3.60$ & $5.35$ & $4.85$\\
\hline\hline
\end{tabular}
\end{center}
\end{table}

The values of $\epsilon^*$, evaluated as reviewed in Section~\ref{subsec:threshold}, are provided in Table~\ref{tab:codewords} for the rate-$1/2$ strongly regular ensembles with VNs in A, S and C forms. The A form exhibits the worst threshold, while the C form achieves the best one. We observe a heavy dependence of the threshold on the VN representation. Note also that the strongly regular C form ensemble achieves a better threshold than the $(3,6)$ regular LDPC ensemble, for which we have $\epsilon^*=0.429$. Next, we searched for a weakly regular ensemble with an optimal mix (from an $\epsilon^*$ viewpoint) of the three VN representations. The problem consists of maximizing $\epsilon^*$ subject to all the CNs being $(7,4)$ Hamming codes, the VNs being length-$7$ SPC codes with A, S or C form and $R=1/2$, where $R$ is given in \eqref{eq:design_rate}. The problem was solved using differential evolution (DE) \cite{price97:differential-evolution}. The optimum weakly regular ensemble (denoted WR in Table~\ref{tab:codewords}) is characterized by a fraction $0.578$ of VNs in A form and a fraction $0.422$ of VNs in S form\footnote{The fact that only the S and A forms are used in the optimal ensemble may be intuitively justified by the fact that the EXIT function of a SPC VN in S form matches tightly the EXIT function of a Hamming CN for values of $I_A$ close to $1$, while the EXIT function of a SPC VN in A form matches tightly the EXIT function of a Hamming CN for values of $I_A$ close to $0$.}. Its threshold ($\epsilon^*=0.481$) is quite larger than that of the $(3,6)$ LDPC ensemble. Remarkably, it has been obtained only by combining different VN representations, without affecting the regularity of the Tanner graph.

Next, we evaluated $G(\alpha)$ for these D-GLDPC ensembles. 
\medskip
\begin{proposition}\label{proposition:SPC_B}
The IO-WEF for the length-$(k+1)$ SPC codes are\footnote{The proof is omitted due to space constraints and can be found in \cite{flanagan09:SPC_VNs}. While the derivation of $B(x,y)$ for the S and A forms is straightforward, for the C form the formula has been obtained using a recursive approach.}:\\
\textbf{S form:}
\begin{align*}
B(x,y) = \frac{1}{2} \left[ (1+y)(1+xy)^{k} + (1-y)(1-xy)^{k} \right]
\end{align*}
\textbf{A form:} ($k_t$ even)
\begin{align*}
B(x,y) & = \frac{1}{2} \left[ (1+xy)^{k} + (1-xy)^{k} \right.\\
\, & \phantom{--------} + \left. y(x+y)^{k} - y(x-y)^{k} \right]
\end{align*}
\textbf{C form:}
\begin{align*}
B(x,y) = 1 + \sum_{i=1}^{k} \sum_{j=1}^{\mathrm{min}(i,k+1-i)} \binom{k+1-i}{j} \binom{i-1}{j-1} x^i y^{2j} \; .
\end{align*}
\end{proposition}
\medskip

\begin{figure}[t]
\begin{center}\includegraphics[%
  width=1.0\columnwidth,
  keepaspectratio]{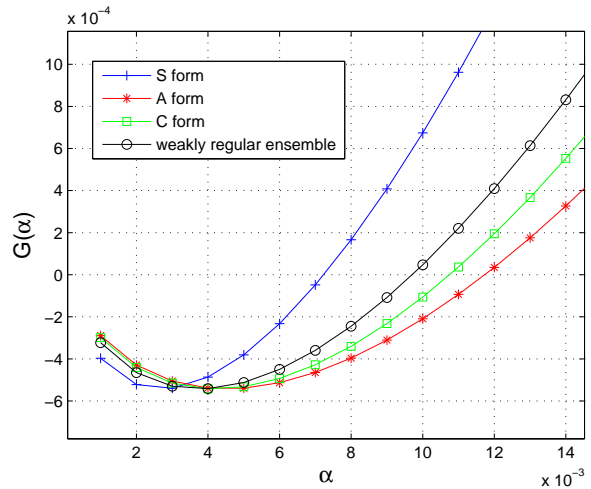}
\end{center}
\caption{Growth rate curves for the rate-$1/2$ strongly and weakly regular D-GLDPC ensembles.}
\label{cap:Hamming_SPC_simulated}
\end{figure}

The plots of growth rate for the three strongly regular ensembles are shown in Fig.~\ref{cap:Hamming_SPC_simulated}. These are evaluated using the method described in Section~\ref{sub:efficient_eval}, which involves solution of a $4 \times 4$ polynomial system. The growth rate for the WR ensemble is also plotted in Fig.~\ref{cap:Hamming_SPC_simulated} based on the solution of a $7 \times 7$ polynomial system following the method described in Section~\ref{sub:efficient_eval}. Note that the same plots can be derived also by implementing \eqref{eq:numerical_evaluation_1_weak} (or \eqref{eq:numerical_evaluation_1_strong}) numerically. However, this approach is characterized by an intrinsic numerical inaccuracy due to the need to quantize the space over which the optimization is performed (finer-grained quantization comes at a price in computational speed, which becomes more pronounced as the optimization space dimensionality increases). The values of $\alpha^{*}$ for the analyzed ensembles  
are reported~in~Table~\ref{tab:codewords}. 

\begin{figure}[t]
\psfrag{q}{\footnotesize{$\epsilon$}}
\psfrag{CER/BER}{\footnotesize{$\mathsf{CER/BER}$}}
\begin{center}
\includegraphics[width=0.70\columnwidth,draft=false,angle=270]{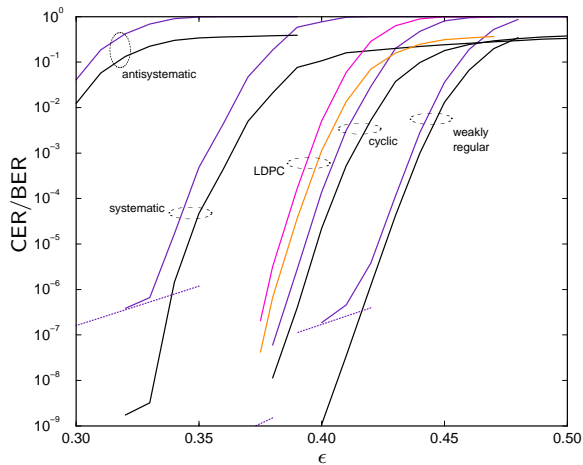}
\end{center}
\caption{Performance over the BEC of $(3000,1500)$ D-GLDPC codes and 
of a $(3000,1500)$ $(3,6)$ regular LDPC code. (CER: codeword error rate. BER: bit error rate. $\epsilon$: BEC erasure probability.)}\label{fig:performance}
\end{figure}

In Fig.~\ref{fig:performance}, the performance curves over the BEC are shown for rate-$1/2$ $N=3000$ D-GLDPC codes from these ensembles. The curves correspond to iterative decoding, with MAP decoding at each node. The four simulated D-GLDPC codes have the same Tanner graph, the only difference being in their SPC VN representations. The Tanner graph was generated using the progressive edge-growth (PEG) algorithm \cite{hu05:peg}, and is composed of $m=500$ degree-$7$ CNs and $n=500$ degree-$7$ VNs. 
For the WR code, 289 VNs are in A form and 211 are in S form (these values target the optimized ensemble found earlier in this section). The performance curve labeled ``LDPC'' in Fig.~\ref{fig:performance} is that of an $N=3000$ $(3,6)$-regular LDPC code generated with the PEG algorithm. The waterfall region of the performance curves reflect the asymptotic thresholds presented in Table~\ref{tab:codewords}, with the WR code exhibiting the best waterfall performance, even if at the price of an error floor at $\mathsf{CER}\simeq 10^{-6}$. We observe how the LDPC code is outperformed in the waterfall region by both the WR and the strongly regular C codes. Again, we observe how a modification in the VN representations can heavily affect the D-GLDPC~code~performance.

An analysis of the error floor was carried out for the simulated D-GLDPC codes by collecting small size stopping sets encountered during the simulations\footnote{It is important to observe that a small-size stopping set (resp. small-weight codeword) collected for any of these D-GLDPC codes represents a small-size stopping set (resp. small-weight codeword) also for the other ones, although with a different size (resp. weight) due to different VN representations.}. Six small-size stopping sets were collected, each one coinciding with a small-weight codeword (labeled `CWD A' to `CWD F'). The weights of such codewords are reported in Table~\ref{tab:codewords}, where $v$ and $c$ denote the number of VNs and CNs involved in the subgraph induced by the codeword, respectively\footnote{The subgraph induced by a codeword is composed of the edges of the Tanner graph carrying a `1' for the given codeword, and the VNs and CNs connected to these edges. Note that, in the subgraph, the edges incident on a VN or CN are associated with a valid local codeword for the node.}. For each code, the smallest among such weights is an estimate of the minimum distance and is reported in bold in Table~\ref{tab:codewords}. We observe that each of these estimates is significantly larger than the corresponding value $\alpha^* n$ for $n=500$, revealing the beneficial effect of a PEG-based construction. On the other hand these estimates are significantly smaller than the value $\alpha^*n=69$ for the $(3,6)$ LDPC code\footnote{From \cite{gallager63:low-density} we have $\alpha^*=0.023$ for the $(3,6)$ LDPC ensemble, so that $\alpha^*n = 0.023 \times 3000=69$ (for the LDPC code we have $n=N=3000$). This value represents a lower bound on the LDPC code minimum distance.}, suggesting that the new codes offer worse error floor properties than the regular LDPC counterparts. 

\begin{figure}[t]
\begin{center}
\includegraphics[width=0.90\columnwidth,draft=false,angle=0]{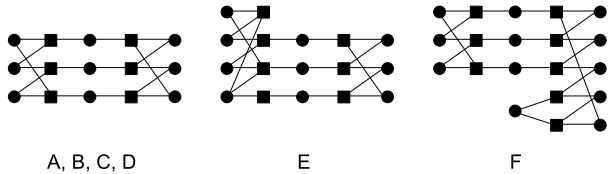}
\end{center}
\caption{Subgraph induced by the codewords listed in Table~\ref{tab:codewords}. (Circles: VNs. Squares: CNs.)}\label{fig:subgraph}
\end{figure}

The estimates of the minimum stopping set sizes were used to calculate a prediction of the error floor. Letting $\theta$ be the minimum stopping set size and assuming a multiplicity one for stopping sets of minimum size, the probability of decoding error $P_e$ fulfills
\begin{align}\label{eq:floor_prediction}
P_e \geq \epsilon^\theta 
\end{align}
as the RHS of \eqref{eq:floor_prediction} is the probability that the starting erasure pattern includes the stopping set of minimum size. As depicted in Fig.~\ref{fig:performance} the lower bound \eqref{eq:floor_prediction} is very tight in the error floor region for the strongly regular S code and WR code. Moreover, it predicts an error floor lower than $\mathsf{CER}=10^{-8}$ for the strongly regular C code which, therefore, exhibits a quite good compromise between waterfall and error floor  performance, while preserving graphical regularity.

The subgraphs induced by the codewords of Table~\ref{tab:codewords} are depicted in Fig.~\ref{fig:subgraph} (the codewords `A' to `D' share the same structure). With the exception of `CWD E', which involves a weight-4 local codeword for one of the Hamming CNs, all the D-GLDPC codewords are associated with local codewords of minimum weight at the nodes, i.e., weight-$3$ codewords at the CNs and weight-$2$ codewords at the VNs. Interestingly, all these subgraphs share a similar structure, composed of a layer of VNs interconnecting two cycles (for `CWD E' one cycle and one structure composed of two overlapping cycles).


\section{Conclusion}
Motivated by the search of new coding schemes with iterative decoding, a class of D-GLDPC codes with Hamming CNs and SPC VNs has been analyzed over the BEC. The asymptotic analysis has been conducted using both EXIT chart and a proposed tool to evaluate the growth rate of the weight distribution. Interesting features recognized from the analysis of a rate-$1/2$ ensemble include the capability of achieving a good compromise between waterfall and error floor performance while preserving graphical regularity, and values of threshold outperforming LDPC~counterparts.

\section*{Acknowledgment}
This work was supported in part by the EC under Seventh FP grant agreement ICT OPTIMIX n.INFSO-ICT-214625. 


\bibliographystyle{IEEEtran}

\end{document}